\documentclass[10pt]{article}

\usepackage[a4paper,height=190mm,width=142mm]{geometry}

\usepackage[utf8]{inputenc}
\usepackage{amsthm}
\usepackage{amssymb}
\usepackage{latexsym}
\usepackage{amsmath}
\usepackage{amsfonts}
\usepackage{mathrsfs}
\usepackage{amstext}
\usepackage{enumitem}
\setitemize{noitemsep,topsep=2pt,parsep=2pt,partopsep=2pt}
\usepackage{mathtools}

\mathtoolsset{showonlyrefs}
\usepackage{stmaryrd}
\usepackage{tocloft}
\usepackage{cite}
\usepackage{graphicx}
\usepackage{xcolor}
\usepackage{color,hyperref}
\definecolor{darkblue}{rgb}{0.0,0.0,0.5}
\hypersetup{
  linkcolor  = black,
  citecolor  = black,
  urlcolor   = black,
  colorlinks = true,
}

\newcommand{\bN}{\mathbb{N}} 
\newcommand{\bE}{\mathbb{E}}
\newcommand{\bC}{\mathbb{C}}
\newcommand{\bZ}{\mathbb{Z}}
\newcommand{\bR}{\mathbb{R}}

\newcommand{\cA}{\mathcal{A}}
\newcommand{\cD}{\mathcal{D}}

\newcommand{\cZ}{\mathcal{Z}}

\newcommand{\cF}{\mathcal{F}}
\newcommand{\cK}{\mathcal{K}}
\newcommand{\cN}{\mathcal{N}}
\newcommand{\cM}{\mathcal{M}}
\newcommand{\sS}{\mathscr{S}}
\newcommand{\sD}{\mathscr{D}}

\newcommand{\rz}{\mathrm{z}}
\newcommand{\rd}{\mathrm{d}}
\newcommand{\re}{\mathrm{e}}

\makeatletter
\newcommand*\botimes{{\mathpalette\botimes@{1.5}}}
\newcommand*\botimes@[2]{\mathbin{\vcenter{\hbox{\scalebox{#2}{\hspace{0.0mm}$\m@th#1\otimes$\hspace{0.0mm}}}}}}
\newcommand*\Cdot{{\mathpalette\Cdot@{.6}}}
\newcommand*\Cdot@[2]{\mathbin{\vcenter{\hbox{\scalebox{#2}{\hspace{0.5mm}$\m@th#1\bullet$\hspace{0.5mm}}}}}}
\makeatother

\newtheorem{thm}{Theorem}[section]
\newtheorem{lem}[thm]{Lemma}
\newtheorem{prop}[thm]{Proposition}

\theoremstyle{dfn}
\newtheorem{dfn}[thm]{Definition}

\newtheorem{Remark}[thm]{Remark}

\numberwithin{equation}{section}
\numberwithin{thm}{section}

\begin{document}
\title{On the $\phi^4_2$ theory on $\bR\times S$}
\author{Azam Jahandideh
\footnote{{\normalsize az.jahan3@gmail.com}
}}

\maketitle

\begin{abstract}
\noindent
We present a construction of the Euclidean two-dimensional $\phi^4$ quantum field theory model on $\mathbb R\times S$ using parabolic stochastic quantization and discuss the Osterwalder-Schrader axioms. Motivated by one-loop string scattering amplitudes, we first construct finite-volume measures on a family of tori with varying geometry. To construct an infinite-volume measure on the cylinder, we establish uniform renormalization and stochastic estimates as the complex structure varies, including the degenerating regime in which the torus approaches an infinite cylinder.
\end{abstract}

\tableofcontents

\section{Introduction}
We revisit the construction of the Euclidean $\phi^4_2$ model and provide a construction of its infinite-volume measure using parabolic stochastic quantization~\cite{parisi1981}. In contrast to our earlier work on the $P(\Phi)_2$ model on the two-sphere~\cite{DDJ}, the present approach relies more systematically on heat-kernel representations to obtain estimates uniform in the complex structure parameter, including the degenerating regime.

Let $\mathfrak{H}=\{\tau=\tau_1+i\tau_2\in\bC:\tau_2>0\}$ be the upper half-plane and $\Sigma_{\tau}:=\bC/(\bZ+\tau\bZ)$ be the two-dimensional torus with complex parameter $\tau\in\cF$, 
where
$$\mathcal{F} = \{ \tau \in \mathfrak{H} \;|\; |\tau| \ge 1,\; -\tfrac12 \le \tau_1 \leq \tfrac12 \}$$
 is the fundamental domain of the modular group $SL(2,\bZ)$. Consider the probability measure on $\sD'(\Sigma_{\tau})$
\begin{equation}\label{eq:measure_mu}
\mu_{\tau}(\rd\phi):=\frac{1}{\cZ_{\tau}}\exp\big(-\,\int_{\Sigma_{\tau}}\,\frac{\lambda}{4}\,:\phi(\rz)^4:\,\rd A_{\tau}(\rz)\big) \,\nu_{\tau}(\rd\phi),
\end{equation}
where $\cZ_{\tau}\in (0,\infty)$ is the partition function, $\lambda\in(0,\infty)$ is the coupling constant, $\nu_{\tau}:= \cN(0,G_{\tau})$ denotes the free field Gaussian measure with covariance 
$G_{\tau}:=(1-\Delta_{\tau})^{-1}$, $:\Cdot:$ refers to the Wick renormalization with respect to the measure $\nu_{\tau}$,  $(-\Delta_{\tau})$ is the Laplace-Beltrami operator on $\Sigma_{\tau}$, and $\rd A_\tau$ is the unnormalized Euclidean area element 
\[
\rd A_\tau=|\frac{i}{2}\,\rd \rz\wedge \rd\bar \rz|=\tau_2\,\rd u\,\rd v,\quad \rz=u\tau+v\in\Sigma_{\tau},\quad\text{with}\quad u,v\in (0,1].
\]
Note that $\Sigma_\tau$ is a flat torus with two fundamental cycles: $u$-cycle of length $|\tau|$ and $v$-cycle of length $1$. Note also that $(\sD'(\Sigma_{\tau}),\mathrm{Borel}(\sD'(\Sigma_{\tau})), \nu_{\tau})$ forms a probability space. The existence of the measure~\eqref{eq:measure_mu} is shown by first introducing a spectrally UV-regularized measure $\mu_{\tau,M}$ and then controlling the limit $M\to \infty$ using Nelson's argument~\cite{Nelson}.

To establish the existence of an infinite-volume measure in the cusp limit $\tau_2\to\infty$, we use parabolic stochastic quantization. Let $\lambda=1$ in Eq.~\eqref{eq:measure_mu}  and denote $Q_{\tau}:=G_\tau^{-1}=(1-\Delta_\tau)$. One writes the parabolic SPDE associated to the measure~\eqref{eq:measure_mu}
\begin{equation}\label{eq:sq_mu}
(\partial_t  + Q_\tau)\,\varPhi_{\tau}(t,\rz)
=- \,:\varPhi^3_\tau(t,\rz):+\sqrt 2\,\xi_\tau(t,\rz),
\quad (t,\rz)\in \bR_+\times \Sigma_\tau,
\end{equation}
where $\xi_\tau$ is the Gaussian white noise with covariance
$$
\bE [\xi_\tau(t,\rz)\,\xi_\tau(t',\rz')]=\delta_{\bR}(t-t')\,\delta_\tau(\rz-\rz').
$$
For $\tau\in\cF$, let $\jmath_\tau\,:\, \mathbb{R}\times S \to \Sigma_\tau$ be the map defined in Def.~\ref{def:covering} and let $\jmath_\tau^*\,:\,\sD'(\Sigma_\tau)\to\sS'(\bR\times S)$ denote the corresponding pull-back of distributions. We denote by $\jmath_\tau^*\sharp\mu_\tau$ the measure on $\sS'(\bR\times S)$, obtained by the push-forward of $\mu_\tau$ with $\jmath_\tau^*$. We shall make use of the SPDE~\eqref{eq:sq_mu} and the intertwining property of the map $\jmath_\tau^*$ in order to prove the existence of accumulation points of the measures $(\jmath_{\tau_n}^*\sharp\mu_{\tau_n})_{n\in\bN_+}$ on $\sS'(\bR\times S)$.

\begin{thm}\label{thm:main}
Let \((\tau_{n})_{n\ge1}\subset\mathcal F\) satisfy \(\mathrm{Im}\tau_{n}\to\infty\). There exists a subsequence \((\tau_{n_k})_{k\ge1}\) such that
$(\jmath_{\tau_{n_k}}^*\sharp\mu_{\tau_{n_k}})$ converges to $\mu$ weakly on \(\sS'(\mathbb R\times S)\).
\end{thm}
\begin{Remark}
In the degenerating limit $\tau_2\to\infty$, the flat torus $\Sigma_\tau$ approaches the infinite cylinder $\mathbb R\times S$. The Gaussian field admits a Kaluza–Klein decomposition along the compact direction, with Fourier modes carrying effective squared masses $\lambda_n^2=1+4\pi^2n^2$, $n\in\bZ$. Thus, the degeneration decompactifies one $\Sigma_\tau$ direction while preserving the discrete spectrum associated with the remaining circle, see~\cite[Ch.~8]{Polchinski} for more details.
\end{Remark}
\begin{Remark}
For the flat metric $g_\tau=|\rd\rz|^2$, the orientation-preserving isometry group of $\Sigma_\tau$ is
\[
\mathrm{Isom}^+(\Sigma_\tau)
\cong\Sigma_\tau\rtimes R_\tau,
\]
where $
R_\tau=\{\omega\in\mathbb C:|\omega|=1,\ \omega\Lambda_\tau=\Lambda_\tau\},
\quad
\Lambda_\tau=\mathbb Z+\tau\mathbb Z$.
The subgroup $R_\tau$ fixing the origin depends on the equivalence class of $\tau\in\cF$ under the modular group. Orientation-reversing isometries exist precisely when the lattice admits a conjugate-linear symmetry
\[
\rz\longmapsto \omega\bar \rz,\quad |\omega|=1,
\qquad \omega\overline{\Lambda_\tau}=\Lambda_\tau.
\]
On the other hand, the isometry group of the limiting cylinder is
\[
\mathrm{Isom}(\mathbb R\times S)
\cong
(\mathbb R\rtimes\bZ_2)
\times(S\rtimes\bZ_2),
\]
generated by translations $
T_a(s,x)=(s+a,x),\quad
P_b(s,x)=(s,x+b)$,
and reflections $\Theta(s,x)=(-s,x)$ and $R(s,x)=(s,-x)$. For every \(\tau\in\mathcal F\), the measure \(\mu_\tau\) is invariant under translations of \(\Sigma_\tau\). These symmetries lift to translations of \(\mathbb R\times S\) using $\jmath_\tau^*$. Consequently, every weak accumulation point \(\mu\) is invariant under \(T_a\) and \(P_b\) for all \(a\in\bR\) and \(b\in S\). Observe that the reflections \(\Theta\) and $R$ on $\bR\times S$ descend to isometries of \(\Sigma_\tau\) if and only if $
2\tau_1\in\bZ$. 
\end{Remark}
\begin{Remark}
The reflection positivity of $\mu_\tau$ can be shown using a Dirac delta approximation. For rectangular tori $\tau=i\tau_2$, the massive Gaussian measure is reflection positive with respect to \(\Theta_\tau[\rz]=[\bar \rz]\). Let \(\rho\in C_c^\infty(\bR^2)\) such that $
\rho\geq0,\quad
\int_{\bR^2}\rho(\rz)\,\rd \rz=1$ with $\mathrm{supp}\rho\subset B_1(0)$,
and $
\rho(-\rz)=\rho(\rz)$ and $\rho(\bar \rz)=\rho(\rz)$.
For sufficiently large \(M\), define its periodization
\[
\rho_{\tau,M}([\rz])
=
M^2\sum_{\lambda\in\mathbb Z+i\tau_2\mathbb Z}
\rho\bigl(M(\rz+\lambda)\bigr).
\]
Then $\mathrm{supp}\rho_{\tau,M}
\subset B_{1/M}([0])$ and $\int_{\Sigma_\tau}\rho_{\tau,M}(\rz)\,\rd A_\tau(\rz)=1$. Note that
$\rho_{\tau,M}$ converges to $\delta_0\in \cD'(\Sigma_\tau)$ by the dominated convergence theorem. Let
\begin{equation}
\begin{aligned}
\Sigma_{\tau,M}^+
=&
\left\{
[x+is]\in\Sigma_\tau:
\frac1M<s<\frac{\tau_2}{2}-\frac1M
\right\},
\\
\Sigma_{\tau,M}^-=&
\left\{[x+is]\in\Sigma_\tau:
-\frac{\tau_2}{2}+\frac1M<s<-\frac1M\right\}.
\end{aligned}
\end{equation}
The regularized covariance \(\widehat G_{\tau,M}=\rho_{\tau,M}\ast G_\tau\ast \rho_{\tau,M}\) preserves reflection positivity for observables supported in $\Sigma_{\tau,M}^+$. The same property holds for the corresponding Wick-ordered interacting measures with the interaction restricted  to $\Sigma_{\tau,M}^+\cup \Sigma_{\tau,M}^-$ and renormalized using $\hat c_{\tau,M}=\hat G_{\tau,M}(0)$ given in Lemma~\ref{lem:hat_counteterm}, and passes to \(\mu_\tau\) upon removing the regularization. One also shows that as $M\to\infty$, the corresponding Gaussian fields and the Wick-ordered interaction converge to the same Gaussian field $X_\tau$ and Wick-ordered interaction $Y_\tau$ given in Prop.~\ref{prop:uv_limit}. Furthermore, using the fact that \(\jmath_\tau^*\) intertwines the torus and cylinder reflections, i.e., $\Theta\circ\jmath^*_\tau=\jmath_\tau^*\circ\Theta_\tau$, where $\Theta(s,x)=(-s,x)$, one infers that the pushforward measures $(\jmath_\tau^*\sharp\mu_\tau)$ are reflection positive on positive-time cylindrical functionals whenever their supports lie within the corresponding torus regions. Consequently, along any rectangular sequence \(\tau_n=i\tau_{2,n}\) with \(\tau_{2,n}\to\infty\), reflection positivity passes to every weak accumulation point \(\mu\) on \(\sS'(\bR\times S)\). 
\end{Remark}
\begin{Remark}
The regularity and integrability of the limiting measures can be verified by adapting the arguments of~\cite{DDJ}, which build on the method of Hairer and Steele~\cite{HairerSteele}. We do not repeat these arguments here.
\end{Remark}
\begin{proof}[Proof of theorem~\ref{thm:main}]
For any sequence $(\tau_n)_{n\in\bN}\subset\cF$ with $\mathrm{Im}\tau_n\to\infty$, the existence of a weakly convergent subsequence of $(\jmath_{\tau_n}^*\sharp\mu_{\tau_n})_{n\in\mathbb N}$ follows from tightness and Prokhorov's theorem. The proof of tightness is presented in Sec.~\ref{sec:tightness} and uses parabolic stochastic quantization combined with a PDE energy estimate.
\end{proof}

In perturbative string theory, scattering amplitudes $\cA$ are obtained by integrating over distributional embedding fields $X$, the insertion points of vertex operators, and the moduli space $\cM$ of worldsheet geometries. For genus one, the moduli space is parametrized by the complex structure parameter $\tau$ modulo the action of the modular group and is equipped with the Poincar\'e measure $\rd\nu=\tau_2^{-2}\,\rd\tau_1\rd\tau_2$. Modular invariance ensures that equivalent representatives of the worldsheet geometry yield the same amplitude $\cA$~\cite{BBK16,Polchinski}. In contrast, in constructive Euclidean QFT, one typically fixes the background geometry rather than integrating over moduli. In heat-kernel and perturbative renormalization, the ultraviolet divergent coefficients are determined by local geometric data; finite renormalized observables can nevertheless depend on global features such as topology~\cite{Va03,BD82,Wald}.

The construction of quantum fields on cylindrical and toroidal geometries is a well-established area of research. The dependence of massless scalar-field partition functions on compactification geometry and modular parameters has been investigated in~\cite{ABB}. Recently, Lorentzian timelike Liouville theory on $\bR\times S$ was constructed through analytic continuation of Euclidean correlation functions obtained from finite-volume torus regularizations~\cite{Ch26}. The construction differs from ours in both the interaction and the underlying Gaussian field; in particular, our Gaussian field is massive. In~\cite[Sec.~IV]{JJM13}, classical fields on $\mathbb R\times\mathbb T^{d-1}$ were treated. For $d=2$, this reduces to the cylinder $\mathbb R\times S$, the same cylindrical geometry as in our setting. 
The cylindrical geometry $\bR\times S$ also plays a role in conformal field theory~\cite{FMS}. The massless Sinh-Gordon model on the infinite cylinder was constructed in~\cite{GGV24}. In a related direction, the $\Phi^4_3$ measure was constructed on arbitrary compact Riemannian three-manifolds using stochastic quantization with universal constants independent of the manifold~\cite{BDFT2}.

The plan of the paper is as follows: 
In Sec.~\ref{sec:uv_limit}, we introduce the regularized $\phi^4_2$ measure $\mu_{\tau,M}$ on $\Sigma_\tau$ with UV cutoff $M\in\bN_+$, and we verify that the measure $\mu_{\tau,M}$ converges weakly, as $M\to\infty$, for fixed $\tau\in\cF$, to the probability measure $\mu_\tau$ formally given by Eq.~\eqref{eq:measure_mu}. We use the Nelson hypercontractivity estimate for this purpose. In Sec.~\ref{sec:stochastic_quantization}, we study the stochastic quantization equations of the measures $\mu_{\tau,M}$ and $\nu_{\tau,M}$ using the Da Prato-Debussche trick. We show local-in-time and global-in-time existence of solutions with fixed $\tau\in\cF$ and $M\in\bN_+$. The covering map $\jmath_\tau$ and its pull-back of distributions $\jmath^*_\tau$ are defined in Sec.~\ref{sec:covering}. In Sec.~\ref{sec:tightness}, we apply the energy technique to establish an a priori bound uniform in \(\tau\in\cF\) and \(M\in\bN_+\). This a priori bound is then used to verify that the sequence of measures $(\jmath_{\tau_n}^*\sharp\mu_{\tau_n})_{n\in\bN_+}$ on $\sS'(\bR\times S)$ admits a weakly convergent subsequence.

\section{Ultraviolet limit}\label{sec:uv_limit}
In order to make sense of the formal expression of Eq.~\eqref{eq:measure_mu} some UV regularization is required. To this end,  we shall use the same choice of UV cut-off as in~\cite{DDJ}. Let $M \in \mathbb{N}_{+}$ and denote $Q_{\tau, M} := (1-\Delta_{\tau})(1 -M^{-2} \Delta_{\tau})^2$. To proceed, we need the following lemma.
\begin{lem}\label{lem:counter-term}
Let $M \in (1,\infty)$ and $G_{\tau,M}=Q_{\tau,M}^{-1}$. Consider
\[
G_{\tau,M}(0)
=\frac{1}{\tau_2}
\sum_{(m,n)\in\mathbb Z^2}
\frac1{
(1+\lambda_{m,n}(\tau))
(1+M^{-2}\lambda_{m,n}(\tau))^2},
\]
where $
\lambda_{m,n}(\tau)
=\frac{4\pi^2}{\tau_2^2}|m-n\tau|^2$. There is $C\in (0,\infty)$ such that 
\begin{equation}
\sup_{\tau\in\cF}\lim_{M\to\infty}\,|\,G_{\tau,M}(0)-\frac{\log M}{2\pi}|<C.
\end{equation}
\end{lem}
\begin{proof}
See appendix~\ref{proof:counteterm}.
\end{proof}
\begin{dfn}
We call $c_{\tau,M}:=G_{\tau,M}(0)$ the counterterm.
\end{dfn}
We write the regularized measure corresponding to the formal measure~\eqref{eq:measure_mu} on $\sD'(\Sigma_\tau)$
\begin{equation}\label{regularized_measure}
\mu_{\tau, M}(\rd\phi)= \frac{1}{\mathcal{Z}_{\tau, M}}\exp\Big(- \int_{\Sigma_{\tau}}\,\mathcal{U}_{\tau,M}(\phi(\rz))\,  \rd A_{\tau}(\rz) \Big) \,\nu_{\tau, M}(\rd\phi),
\end{equation}
where $\mathcal{U}_{\tau,M}(\phi(\rz)):=1/4\,\phi(\rz)^{4} -3/2\, c_{\tau, M}\,\phi(\rz)^{2}+3/4\,c_{\tau,M}^2$, $\nu_{\tau, M}=\cN(0,G_{\tau,M})$ is the Gaussian measure with mean zero and covariance $G_{\tau,M}=(1-\Delta_\tau)^{-1}(1-M^{-2}\Delta_\tau)^{-2}$.
\begin{lem}\label{lem:mu_measure_well_defined}
For all $\tau\in\cF$ and $M\in\mathbb{N}_+$, the measure $\mu_{\tau,M}$ is well-defined and both $\nu_{\tau,M}$ and $\mu_{\tau,M}$ are concentrated on $L^1_2(\Sigma_\tau)\subset\sD'(\Sigma_\tau)$.
\end{lem}
\begin{proof}
Lemmas~\ref{lem:sigma_1}~and~\ref{lem:lp-l2} imply $
\nu_{\tau,M}\big(L_2^1(\Sigma_\tau)\cap L_4(\Sigma_\tau)\big)=1$.
Moreover, $
\mathcal U_{\tau,M}(\phi)\geq -3/2\,c_{\tau,M}^2$. Consequently,
\begin{equation}
0<
\exp\left(-\int_{\Sigma_\tau}
\mathcal U_{\tau,M}(\phi(\rz))\,\rd A_\tau(\rz)\right)\le \re^{3/2\tau_2\,c_{\tau,M}^2}
\end{equation}
for \(\phi\in L_4(\Sigma_\tau)\). The density is therefore bounded, positive and measurable. Finally, Jensen's inequality and Fubini's theorem yield
$\cZ_{\tau,M}\ge 1$
because the fourth Wick-ordered field has zero expectation. Hence $
1\leq \mathcal Z_{\tau,M}\le \re^{3/2\tau_2\,c_{\tau,M}^2}<\infty$. The measure \(\mu_{\tau,M}\) is thus well-defined, absolutely continuous with respect to \(\nu_{\tau,M}\), and is concentrated on \(L_2^1(\Sigma_\tau)\).
\end{proof}
Let $K_{\tau,M}:=(1-M^{-2}\Delta_\tau)^{-1}$. To proceed, define the random Gaussian variables $X_{\tau}$ and $X_{\tau,M}=K_{\tau,M}X_{\tau}$ on $(\sD'(\Sigma_{\tau}),\mathrm{Borel}(\sD'(\Sigma_{\tau})), \nu_{\tau})$ such that $\mathrm{Law}(X_{\tau})= \nu_{\tau}$ and $\mathrm{Law}(X_{\tau,M})=\nu_{\tau,M}$. One has in distributional sense
$$\bE\,[X_{\tau}\otimes X_{\tau}] = G_{\tau},
\quad
\bE\,[X_{\tau,M}\otimes X_{\tau,M} ]= G_{\tau,M}.$$
From Lemma~\ref{lem:sigma_1} one infers that $X_{\tau,M}\in L^{2-\kappa}_2(\Sigma_\tau)\subset L_4(\Sigma_\tau)$ almost surely. For every bounded and continuous $F\,:\, \sD'(\Sigma_{\tau})\to\bR$, we have
\begin{equation}
\begin{gathered}
\int F(\phi)\,\nu_{\tau,M}(\rd\phi)=\bE\,F(X_{\tau,M}),
\quad
 \int F(\phi)\, \mu_{\tau,M}(\rd\phi)
 =
 \frac{\bE\,F(X_{\tau,M})\exp(-\,Y_{\tau,M})}{\bE\exp(-\,Y_{\tau,M})}\,,
 \end{gathered}
\end{equation}
where $Y_{\tau,M}:=\frac{1}{4}\int_{\Sigma_\tau}\,X_{\tau,M}^{:4:}\,\rd A_\tau$.

\begin{prop}\label{prop:uv_limit}
Let $\tau\in\cF$. There exist random variables $X_\tau$ and $Y_\tau$ such that $\bE\,\exp(-Y_\tau)<\infty$ and for every bounded and continuous $F\,:\,\sD'(\Sigma_\tau)\to \bR$ and all $p\in (0,\infty)$ it holds that
\begin{equation}
 \lim_{M\to\infty} \bE\, [F(X_{\tau,M}) \,\exp(-p\,Y_{\tau,M}) ]= \bE\,[ F(X_\tau)\, \exp(-p\,Y_\tau)].
\end{equation}
\end{prop}
\begin{proof}
One can show that  the sequence $(F(X_{\tau,M}) \exp(-\,Y_{\tau,M}))_{M\in\bN_+}$
converges in probability to $F(X_\tau) \exp(-\,Y_\tau)$ using Lemma~\ref{lem:stochastic_X_uniform} and Lemma~\ref{lem:stochastic_Y}. Moreover, there exist constants $C,\,c\in (0,\infty)$ such that for all $M\in (1,\infty)$ and all $t> 0$ 
\begin{equation}
 \mathbb{P}(-\,Y_{\tau,M}> t) \leq C\,\exp(-\exp(c\,\sqrt t)).
\end{equation}
This tail estimate implies uniform integrability of the sequence
$(F(X_{\tau,M}) \,\exp(-\,Y_{\tau,M}))_{M\in\bN_+}$.
The Vitali theorem concludes the statement. See~\cite[Prop.~2.7]{DDJ} for more details. 
\end{proof}

\section{Stochastic quantization}\label{sec:stochastic_quantization}
The parabolic stochastic quantization corresponding to $\mu_{\tau,M}$ is given by
\begin{equation}\label{eq:stoc_sig}
(\partial_t+Q_{\tau,M})\,\varPhi_{\tau,M}(t,\rz)=\sqrt{2}\,\xi_{\tau}(t,\rz)-\varPhi^3_{\tau,M}(t,\rz)+3\,c_{\tau,M}\,\varPhi_{\tau,M}(t,\rz),
\quad (t,\rz)\in \bR\times \Sigma_\tau.
\end{equation}
We aim to prove that the measure $\mu_{\tau,M}(\rd\phi)$ is invariant under the dynamics generated by Eq.~\eqref{eq:stoc_sig}, see Lemma~\ref{lem:meaures_sig_1}. This implies that if $\varPhi_{\tau,M}$ is a stationary solution of Eq.~\eqref{eq:stoc_sig} with initial condition $\varPhi_{\tau,M}(0,\Cdot) = \phi_{\tau,M}$ distributed according to the measure $\mu_{\tau,M}(\rd\phi)$, then for all $t\in \bR_+$ the random field $\varPhi_{\tau,M}(t,\Cdot)$ is also distributed according to $\mu_{\tau,M}(\rd\phi)$. To this end, we utilize the Da Prato-Debussche decomposition~\cite{daprato2003}. Throughout this section, we fix a stochastic basis $(\Omega, \cF,\{\cF_t\}_{t\in[0,\infty)},\mathbb{P})$ and we consider the cylindrical Wiener process $(W_\tau(t,\Cdot))_{t\geq 0}$ on $L_2(\Sigma_\tau)$. Consider
\begin{equation}\label{eq:weak_free}
(\partial_t+Q_{\tau,M})Z_{\tau,M}(t,\rz)=\sqrt{2}\,\xi_\tau(t,\rz),\quad (t,\rz)\in\bR\times \Sigma_\tau\,,
\end{equation}
where $\bE\xi_\tau(t,\rz)\,\xi_\tau(t',\rz')=\delta_{\bR}(t-t')\,\delta_{\tau}(\rz-\rz')$.
\begin{lem}\label{lem:free_solution}
The unique mild solution $Z_{\tau,M}\in C([0,\infty),L^1_2(\Sigma_\tau))$
of Eq.~\eqref{eq:weak_free} with the initial condition $Z_{\tau,M}(0,\Cdot)=z_{\tau,M}\in  L^1_2(\Sigma_\tau)$, which is independent of $W_\tau$, is given by
\begin{equation}\label{eq:mild_free}
 Z_{\tau,M}(t,\Cdot) = \re^{-t\,Q_{\tau,M}} z_{\tau,M}+ \sqrt{2}\,\int_0^t\re^{-(t-s)\,Q_{\tau,M}}\,\rd W_\tau(s,\Cdot)\,.
\end{equation}
\end{lem}
\begin{proof}
Let $T>0$. Fix \(\tau\in\mathcal F\), \(M>1\), and choose \(0<\alpha<1/6\). Then
\begin{multline}
\int_0^T s^{-2\alpha}
\|\re^{-sQ_{\tau,M}}\|_{\mathrm{HS}(L_2,L_2^1)}^2\,\rd s
\\
\le
\Gamma(1-2\alpha)
\sum_{m,n\in\mathbb Z}
(1+\lambda_{m,n}(\tau))
\left[
2(1+\lambda_{m,n}(\tau))
(1+M^{-2}\lambda_{m,n}(\tau))^2
\right]^{2\alpha-1}
<\infty.
\end{multline}
The Burkholder-Davis-Gundy theorem~\cite[Thm.~5.2.4]{DaZab96} and the fact that 
the semigroup $t\mapsto \re^{-tQ_{\tau,M}}$ is uniformly bounded in the operator norm on $L_2^1(\Sigma_\tau)$ on any time interval conclude the proof.
\end{proof}
One writes the stochastic quantization equation associated to $\mu_{\tau,M}$
\begin{equation}\label{eq:weak_interacting}
\rd\varPhi_{\tau,M}(t,\Cdot)
=\sqrt2\,\rd W_\tau(t,\Cdot)
-\left(Q_{\tau,M}\,\varPhi_{\tau,M}(t,\Cdot)
+\varPhi_{\tau,M}^3(t,\Cdot)-3\,c_{\tau,M}\,\varPhi_{\tau,M}(t,\Cdot)\right)\,\rd t.
 \end{equation}
This equation is well-defined for fixed $\tau\in\cF$ and $M\in\bN_+$. 
\begin{lem}\label{lem:mild_solution_interacting}
There exists a unique mild solution $\varPhi_{\tau,M}\in C([0,\infty),L^1_2(\Sigma_\tau))$ to Eq.~\eqref{eq:weak_interacting}, with initial condition $\varPhi_{\tau,M}(0,\Cdot)=\phi_{\tau,M}\in L^1_2(\Sigma_\tau)$ independent of $W_\tau$.
\end{lem}

\begin{Remark}\label{rem:initial_date}
Note that for all $\tau\in\cF$ and $M\in\bN_+$, both initial conditions $z_{\tau,M},\phi_{\tau,M}\in L^1_2(\Sigma_\tau)$ are independent of $W_\tau$ and satisfy $\mathrm{Law}(z_{\tau,M},\phi_{\tau,M})
=\nu_{\tau,M}\times\mu_{\tau,M}$.
\end{Remark}
Eq.~\eqref{eq:weak_interacting} becomes singular in the limit $M\to\infty$. Invoking the Da Prato-Debussche trick~\cite{daprato2003}, we write, for all $t\in [0,\infty)$, the random field $\varPhi_{\tau,M}(t,\Cdot)$ as
\begin{equation}\label{eq:DDD}
\varPhi_{\tau,M}(t,\Cdot) = \varPsi_{\tau,M}(t,\Cdot) + Z_{\tau,M}(t,\Cdot),
\end{equation}
where $Z_{\tau,M}(t,\Cdot)$ is given in Eq.~\eqref{eq:weak_free} and $\varPsi_{\tau,M}(t,\Cdot)$ is the mild solution to the following PDE
\begin{equation}\label{eq:varPsi_pde}
(\partial_t+Q_{\tau,M})\varPsi_{\tau,M}
=-\varPsi_{\tau,M}^3
-3\,\varPsi_{\tau,M}^2Z_{\tau,M}
-3\,\varPsi_{\tau,M}\,Z_{\tau,M}^{:2:}
-Z_{\tau,M}^{:3:}\,,
\end{equation}
where $
Z_{\tau,M}^{:2:}
=Z_{\tau,M}^2-c_{\tau,M}$ and $Z_{\tau,M}^{:3:}
=Z_{\tau,M}^3-3\,c_{\tau,M}\,Z_{\tau,M}$.
\begin{proof}[Proof of Lemma~\ref{lem:mild_solution_interacting}]
 By Lemma~\ref{lem:free_solution}, the mild solution $Z_{\tau,M}(t,\Cdot)\in C([0,\infty), L^1_2(\Sigma_\tau))$ exists and is unique. For every initial condition $\psi_{\tau,M}\in L^1_2(\Sigma_\tau)$, the existence and uniqueness of the global-in-time mild solution to Eq.~\eqref{eq:varPsi_pde} follows from Proposition~\ref{prop:global_in_time_existence}. This completes the proof.
\end{proof}
Let
\begin{equation}\label{eq:non-linear-Psi}
F( \varPsi_{\tau,M}(s,\Cdot)):=-\varPsi_{\tau,M}^3
-3\,\varPsi_{\tau,M}^2\,Z_{\tau,M}
-3\,\varPsi_{\tau,M}\,Z_{\tau,M}^{:2:}
-Z_{\tau,M}^{:3:}
\end{equation}
and extend this definition from $s\mapsto \varPsi_{\tau,M}(s,\Cdot)$ to arbitrary $\varPsi\in C([0,T] ; L^1_2(\Sigma_\tau))$.
\begin{lem}\label{lem:lipschitz}
Consider the mapping $\cK$ defined by
\begin{equation}
\cK(\varPsi)(t,\Cdot):= \re^{-t\,Q_{\tau,M}}\,\psi_{\tau,M} + \Gamma(\varPsi)(t,\Cdot),
\quad
\Gamma(\varPsi)(t,\Cdot) = \int_0^t\,\re^{-(t-s)\,Q_{\tau,M}}\,F(\varPsi(s,\Cdot))\,\rd s,
\end{equation}
where $\psi_{\tau,M}:=\phi_{\tau,M}- z_{\tau,M}\in L^1_2(\Sigma_\tau)$. It holds that
\begin{equation}
\cK\,:\,C([0,T] ;L^1_2(\Sigma_\tau)) \to C([0,T] ;L^1_2(\Sigma_\tau))
\end{equation}
and there exists a constant $C\in (0,\infty)$ such that for every $\varPsi_1,\varPsi_2\in C([0,T];L^1_2(\Sigma_\tau))$, one has
\begin{equation}\label{eq:lip_bound_non_lineararity}
\|\cK(\varPsi_1) - \cK(\varPsi_2)\|_{C([0,T]; L^1_2(\Sigma_\tau))}
\leq C\,T^{\,5/6}\,D(\varPsi_1,\varPsi_2)\,\,\|\varPsi_1 -\varPsi_2\|_{C([0,T];L^1_2(\Sigma_\tau))},
\end{equation}
where 
\begin{equation}
D(\varPsi_1,\varPsi_2):=\Big[2+ \sum_{l=1}^{2}\,\left(\|\varPsi_1\|^{l}_{C([0,T];L^1_2(\Sigma_\tau))} + \|\varPsi_2\|^{l}_{C([0,T];L^1_2(\Sigma_\tau))}\right)\Big]
\end{equation}
and the constant \(C\) may depend on the realization \(\omega\in\Omega\).
\end{lem}
\begin{proof}
The proof follows from H\"older's inequality, the Sobolev embeddings $L_2^1(\Sigma_\tau)\rightarrow L_p(\Sigma_\tau)$ given in Lemma~\ref{lem:lp-l2}, the fact that $F\,:\,L^1_2(\Sigma_\tau)\to L_2(\Sigma_\tau)$ is locally Lipschitz continuous, the regularity \(Z_{\tau,M}\in C([0,T];L_2^{2-\kappa}(\Sigma_\tau))\) for \(\kappa\in(0,1)\), and the semigroup estimate $\|\re^{-tQ_{\tau,M}}\|_{L^2(\Sigma_\tau)\to L_2^1(\Sigma_\tau)}
\le C_M \,t^{-1/6},\quad t>0$.
\end{proof}
\begin{thm}\label{thm:mild_solution_non-linear}
There exists $T>0$ depending on $\tau,M,\omega$, and $\|\psi_{\tau,M}\|_{L^1_2(\Sigma_\tau)}$ such that Eq.~\eqref{eq:varPsi_pde} has a unique mild solution $\varPsi_{\tau,M}\in C([0,T] ;L^1_2(\Sigma_\tau))$.
\end{thm}
\begin{proof}
The proof is an application of the standard contraction mapping argument combined with the preceding lemma.
\end{proof}
\begin{lem}\label{lem:mild_regularity_time_space}
It holds that
\begin{equation}
 \varPsi_{\tau,M}\in C([0,\infty), L^1_2(\Sigma_\tau))\cap C((0,\infty),L^3_2(\Sigma_\tau))\cap C^1((0,\infty),L^{-3}_2(\Sigma_\tau))
\end{equation}
and the following equality $
\partial_t\varPsi_{\tau,M} 
= -Q_{\tau,M}\varPsi_{\tau,M}-\varPsi_{\tau,M}^3
-3\varPsi_{\tau,M}^2Z_{\tau,M}
-3\varPsi_{\tau,M}\,Z_{\tau,M}^{:2:}
-Z_{\tau,M}^{:3:}$ is satisfied in $C((0,\infty),L^{-3}_2(\Sigma_\tau))$. 
\end{lem}
\begin{proof}
By Lemma~\ref{lem:free_solution} and Sobolev embeddings, the nonlinear map $F$ given in~\eqref{eq:non-linear-Psi} is locally Lipschitz from both $L_2^1(\Sigma_\tau)$ and $L_2^3(\Sigma_\tau)$ into $L_2(\Sigma_\tau)$ for almost every realization of \(Z_{\tau,M}\). The smoothing estimate $
\|\re^{-tQ_{\tau,M}}f\|_{L_2^3(\Sigma_\tau)}
\leq C_{M}\,t^{-1/3}\|f\|_{L_2^1(\Sigma_\tau)}
$
and the dominated convergence theorem applied to the mild formulation yield $\varPsi_{\tau,M}\in C((0,\infty),L_2^3(\Sigma_\tau))$. Differentiating the mild formulation in $L_2^{-3}(\Sigma_\tau)$ gives the asserted time regularity and evolution equation.
\end{proof}
We aim to establish the existence of a global-in-time mild solution to Eq.~\eqref{eq:varPsi_pde} using the energy method  in  $L^1_2(\Sigma_\tau)$. The proof relies on the fact that for all $m \in \{1,2,3\}$, $p\in[2,\infty)$, and all $\tau\in\cF$ and $M\in\bN_+$, the renormalized free fields $Z_{\tau,M}^{:m:}(t,\Cdot)$, and their derivatives $\nabla_\tau\,Z_{\tau,M}^{:m:}(t,\Cdot)$ are uniformly bounded  in $L_p(\Sigma_\tau)$ over $t\in [0,T]$. This follows from Lemma~\ref{lem:sigma_1} 
and the polynomial representation of the Wick powers in $Z_{\tau,M}$ for fixed $M\in\bN_+$.
\begin{prop}\label{prop:global_in_time_existence}
For all $\tau\in\cF$ and $M\in\bN_+$, there exist $p\in[2,\infty)$ and $C\in(0,\infty)$ such that for all $t\in (0,\infty)$ it holds that
\begin{multline}
\begin{aligned}
\partial_t\|\varPsi_{\tau,M}(t,\Cdot)\|^2_{L^1_2(\Sigma_\tau)} 
&
+M^{-4}\,\|\varPsi_{\tau,M}(t,\Cdot)\|^2_{L^4_2(\Sigma_\tau)} 
+\|\varPsi_{\tau,M}(t,\Cdot)\|^4_{L_4(\Sigma_\tau)}
\\&
+3\,\int_{\Sigma_\tau}|\nabla_\tau\varPsi_{\tau,M}(t,\rz)|^2\,(\varPsi_{\tau,M})(t,\rz)^{2}\,\rd A_\tau(\rz)
\\&
\leq C\sum_{k=0}^{3}\,(\|Z^{:k:}_{\tau,M}(t,\Cdot)\|^p_{L_p(\Sigma_\tau)} +\|\nabla_\tau Z_{\tau,M}^{:k:}(t,\Cdot)\|^p_{L_p(\Sigma_\tau)}).
\end{aligned}
\end{multline}
\end{prop}
\begin{proof}
Let $\varPsi_{\tau,M}(t,\Cdot) := \varPsi(\Cdot)$ and $Z^{:m:}_{\tau,M}(t,\Cdot) := Z^{:m:}(\Cdot)$.
Multiplying both sides of Eq.~\eqref{eq:varPsi_pde} by $(1-\Delta_\tau)\varPsi$ and integrating over $\Sigma_\tau$, we obtain the following terms
\begin{itemize}

\item[(A)] $\langle (1-\Delta_\tau)\varPsi, \partial_{t}\varPsi\rangle_{L_2(\Sigma_\tau)}= \frac{1}{2} \partial_t\,\|\varPsi\|^2_{L^1_2(\Sigma_\tau)}$\,.

\item[(B)] $\langle (1-\Delta_\tau)\varPsi, Q_{\tau,M}\,\varPsi \rangle_{L_2(\Sigma_\tau)} \geq M^{-4}\,\|\varPsi \|^2_{L^4_2(\Sigma_\tau)}$\,.

\item[(C)] $\langle (1-\Delta_\tau)\varPsi,  \varPsi^{3}\rangle_{L_2(\Sigma_\tau)} = \|\varPsi\|^4_{L_4(\Sigma_\tau)}
+3\,\int_{\Sigma_\tau} |\nabla_\tau\varPsi(\rz)|^2  |\varPsi(\rz) |^{2}\,\rd A_\tau(\rz)$\,.

\item[(D)] $\langle (1-\Delta_\tau)\varPsi,
-3\,\varPsi^2\,Z-3\,\varPsi\,Z^{:2:}-Z^{:3:} \rangle_{L_2(\Sigma_\tau)} $\,.

\end{itemize}
We estimate the terms in Item~(D) in terms of Items~(B) and~(C). To this end, we shall repeatedly use the fact that $(1-\Delta_\tau)$ is a positive and self-adjoint operator, integration by parts, and H\"older's and Young's inequalities. One concludes using Lemma~\ref{lem:sigma_1} and Remark~\ref{rem:bessel_sobolev_psi}.
\end{proof}
Note that the preceding proposition provides an a priori $L_2^1(\Sigma_\tau)$ estimate for $\varPsi_{\tau,M}(t,\Cdot)$. This implies the existence of a global-in-time mild solution to Eq.~\eqref{eq:varPsi_pde} in the space $L_2^1(\Sigma_\tau)$.

\begin{lem}\label{lem:meaures_sig_1}
For $t\geq 0$ and all $\tau\in\cF$ and $M\in\bN_+$: (A) $\mathrm{Law}(Z_{\tau,M}(t,\Cdot))=\mathrm{Law}(z_{\tau,M})=\nu_{\tau,M}$ and~(B) $\mathrm{Law}(\varPhi_{\tau,M}(t,\Cdot))=\mathrm{Law}(\phi_{\tau,M})=\mu_{\tau,M}$
on $\sD'(\Sigma_\tau)$. 
\end{lem}
\begin{proof}
For all $\tau\in\cF$ and $M\in\bN_+$, consider the linear operator $Q_t\,:\, L^1_2(\Sigma_\tau)\to L^1_2(\Sigma_\tau)$ defined by $Q_t \phi:=2\,\int_0^t\,\re^{-2s\,Q_{\tau,M}}\,\phi\,\rd s$ for $\phi\in L^1_2(\Sigma_\tau)$. Note that $\sup_{t\geq 0}\,\mathrm{Tr}\,Q_t<\infty$. Moreover, $\cN(0,Q_{\infty})=\cN(0,G_{\tau,M})=\nu_{\tau,M}$. This proves Item~(A) thanks to Remark~\ref{rem:initial_date}. For Item (B), $Q_{\tau,M}$ is positive and self-adjoint with trace-class inverse, and the nonlinear drift is the negative gradient of the interaction functional, which is bounded from below by Lemma~\ref{lem:mu_measure_well_defined}. Furthermore, by Lemma~\ref{lem:free_solution}, the Ornstein-Uhlenbeck process $Z_{\tau,M}(t,\Cdot)\in C([0,\infty);L^1_2(\Sigma_\tau)) $ and from the proof of Lemma~\ref{lem:lipschitz}, the map
 $F\,:\,L^1_2(\Sigma_\tau)\to L_2(\Sigma_\tau)$ is locally Lipschitz continuous. Together with Lemma~\ref{lem:mild_solution_interacting}, these properties yield invariance of $\mu_{\tau,M}$ by~\cite[Hypothesis~8.4~and~Thm.~8.6.3]{DaZab96}.
\end{proof}

\begin{Remark}\label{rem:meaures_cyl_1}
Combining the pushforward property of probability laws with the preceding results, it follows that for all $\tau\in\cF$ and $M\in\bN_+$, and all $t\geq 0$, $\mathrm{Law}(\jmath_\tau^*Z_{\tau,M}(t,\Cdot))=\mathrm{Law}(\jmath_\tau^*z_{\tau,M})=\jmath^*_\tau\sharp\nu_{\tau,M}$, and $\mathrm{Law}(\jmath_\tau^*\varPhi_{\tau,M}(t,\Cdot))=\mathrm{Law}(\jmath_\tau^*\phi_{\tau,M})=\jmath^*_\tau\sharp\mu_{\tau,M}$
on $\sS'(\bR\times S)$. 
\end{Remark}
\section{Covering map}\label{sec:covering}
In this section, we aim to establish a relation between the geometry of the flat torus $\Sigma_\tau$ and that of the infinite cylinder $\mathbb R\times S$, where $S=\mathbb R/\mathbb Z$. We introduce a covering map that identifies the torus with a quotient of the cylinder. As $\tau_2\to\infty$, with $\tau_1$ bounded, the flat torus converges locally to the infinite cylinder. This representation allows us to formulate estimates on the cylinder using weighted function spaces, with constants that remain uniform as the torus degenerates.

\begin{dfn}\label{def:covering}
Let \(\rz\in\mathbb C\) and \(x\in\mathbb R\). We define equivalence classes \([\rz]\) and \([x]\) 
\begin{equation}
\begin{gathered}
[\rz]=\rz+(\mathbb Z+\tau\mathbb Z)
=\{\rz+m+n\tau:m,n\in\mathbb Z\}
\in\Sigma_\tau,
\quad
[x]=x+\mathbb Z
=\{x+n:n\in\mathbb Z\}
\in S.
\end{gathered}
\end{equation}
Let $s=\tau_2u$ and $x=v+\tau_1u$. Define
\[
\jmath_\tau\,:\,\mathbb R\times S\longrightarrow\Sigma_\tau,
\qquad
\jmath_\tau(s,[x])=[x+is]=[\rz]=[\tau u+v].
\]
It satisfies
\[
\jmath_\tau(s+\tau_2,[x+\tau_1])
=\jmath_\tau(s,[x])=[\rz]=[\tau u+v].
\]
Locally, $\rz=\tau u+v$. Therefore,
$$
\jmath_\tau^*g_\tau=\jmath_\tau^*|\rd \rz|^2= \rd s^2+\rd x^2,
\quad \jmath_\tau^*\rd A_\tau=\rd s\,\rd x\,.
$$
Then for all $f\in C^{\infty}(\Sigma_\tau)$ one has
\begin{equation}\label{eq:intertwining}
\jmath_\tau^*(\Delta_\tau f)
=(\partial_s^2+\partial_x^2)(\jmath_\tau^*f):=\Delta_{\mathrm{cyl}}\,(\jmath_\tau^*f),
\qquad
\jmath_\tau^*\rd A_\tau=\rd s\,\rd x.
\end{equation}
\end{dfn}
\begin{Remark}\label{rem:weight_2}
Let $\tau\in\cF$ and $f\in C(\Sigma_\tau)$. Then $\jmath_\tau^* f=f\circ \jmath_\tau\in C(\bR\times S)$. Every smooth function $ g\in C^\infty(\bR\times S)$ satisfying
$g(s+\tau_2,x+\tau_1)=g(s,x)$ descends to a unique smooth function on \(\Sigma_\tau\). It holds
\begin{equation}
\int_{(0,\tau_2]\times S}\,\jmath^*_\tau f(s,x)\,\rd s\,\rd x=\int_{(0,\tau_2]\times S}\,f\circ \jmath_\tau(s,x)\,\rd s\,\rd x
=\int_{\Sigma_\tau}\,f(\rz)\,\rd A_\tau(\rz).
\end{equation}
Note that as $\tau_2\to\infty$, the longitudinal direction becomes noncompact. Moreover, \(\jmath_\tau^*f\) is periodic in the longitudinal direction, up to the twist:
\[
(\jmath_\tau^*f)(s+\tau_2,x+\tau_1)
=(\jmath_\tau^*f)(s,x).
\]
Consequently, integrating over the entire infinite cylinder counts infinitely many copies of the torus. 
On the cylinder $\bR\times S$, consider a weight $w(s)=(1+as^2)^{-b/2}$ for some $b>2$ and $a\in (0,1)$ sufficiently small and use its periodization, i.e., $w_\tau$ on $\Sigma_\tau$
\[
w_\tau(\jmath_\tau(s,[x]))
=
\sum_{k\in\mathbb Z}
\left(1+a(s+k\tau_2)^2\right)^{-b/2},
\quad \tau\in\cF.
\]
For \(b>2\), the series converges smoothly and satisfies
\[
w_\tau(\jmath_\tau(s+\tau_2,[x+\tau_1]))
=w_\tau(\jmath_\tau(s,[x])).
\]
For every \(\tau\in\mathcal F\)
\[
\int_{\Sigma_\tau} w_\tau\,\rd A_\tau=
\int_0^{\tau_2}\sum_{k\in\mathbb Z}w(s+k\tau_2)\,\rd s
=\sum_{k\in\mathbb Z}\int_{k\tau_2}^{(k+1)\tau_2}w(s)\,\rd s
=\int_{\mathbb R}\,w(s)\,\rd s,
\]
where we used the fact that the circle has unit length. This gives a uniform weighted-volume estimate compatible with the cylindrical limit.
\end{Remark}
\begin{Remark}\label{rem:weight}
For the polynomial weight $w(s)=(1+as^2)^{-b/2}$, with $b>2$ and sufficiently small $a\in (0,1)$, we have
\[
|\nabla^jw|\le C_b\sqrt a\,w,
\qquad 1\le j\le6.
\]
Moreover, by H\"older's inequality,
\[
\|\Psi\|_{L_2(w^{1/2})}
\le \left(\int_{\mathbb R\times S}w\,\rd s\,\rd x\right)^{1/4}
\|\Psi\|_{L_4(w^{1/4})} \leq C\,\|\Psi\|_{L_4(w^{1/4})},
\]
where $C\in (0,\infty)$ depends only on \(a,b\), and is independent of \(\tau,M,\Psi\).
\end{Remark}

\section{Tightness}\label{sec:tightness}
Using Eq.~\eqref{eq:intertwining}, one rewrites Eq.~\eqref{eq:varPsi_pde} on $\bR\times S$ 
\begin{multline}\label{eq:eq_cylider_1}
(\partial_t+
(1-\Delta_{\mathrm{cyl}})
(1-M^{-2}\Delta_{\mathrm{cyl}})^{2})\,\jmath_\tau^*\varPsi_{\tau,M}
\\
=-(\jmath_\tau^*\varPsi_{\tau,M})^3
-3\,(\jmath_\tau^*\varPsi_{\tau,M})^2
(\jmath_\tau^*Z_{\tau,M})
-3\,(\jmath_\tau^*\varPsi_{\tau,M})
(\jmath_\tau^*Z_{\tau,M}^{:2:})
-\jmath_\tau^*Z_{\tau,M}^{:3:}.
\end{multline}
\begin{prop}\label{prop:energy_method_plane}
There exist $\kappa\in(0,1/3)$, $C\in(0,\infty)$, $p\in[1,\infty)$ such that for all $t\in(0,\infty)$, $\tau\in\cF$, and $M\in\bN_+$, it holds
\begin{equation}
\,\partial_t\|\jmath_\tau^* \varPsi_{\tau,M}(t,\Cdot)\|^2_{L_2(\bR\times S,w^{1/2})}
 +
 \|\jmath_\tau^* \varPsi_{\tau,M}(t,\Cdot)\|^4_{L_4(\bR\times S,w^{1/4})}
\leq C \sum_{k=0}^{3}\,
\|\jmath_\tau^*Z_{\tau,M}^{:k:}(t,\Cdot)\|^p_{L_p^{-\kappa}(\bR\times S, w^{1/p})}\,.
\end{equation}
\end{prop}
\begin{proof}
Let $Q_{\mathrm{cyl},M}:=(1-\Delta_{\mathrm{cyl}})
(1-M^{-2}\Delta_{\mathrm{cyl}})^{2}$. The proof is an application of the energy method.
After multiplying both sides of Eq.~\eqref{eq:eq_cylider_1} by $w\jmath_\tau^*\varPsi_{\tau,M}$, integrating over $\bR\times S$, one obtains the coercivity estimate
\[
\left\langle\Psi,Q_{\mathrm{cyl},M}\Psi\right\rangle_{L_2(\bR\times S,w^{1/2})}
\ge\frac12
\|\Psi\|_{L_2(\bR\times S,w^{1/2})}^2+\frac{1}{2}\|\nabla\Psi\|_{L_2(\bR\times S,w^{1/2})}^2.\] 
This estimate follows from integration by parts, the Cauchy-Schwarz inequality, and the Young inequality. Note that the derivatives of the weight are controlled using Remark~\ref{rem:weight}. Choosing the parameter in the weight sufficiently small allows the resulting error terms to be absorbed uniformly in $M\geq 1$. The conclusion then follows from Lemma~\ref{lem:Z_Psi_4}.
\end{proof}
To proceed, we establish tightness of the family $(\jmath_{\tau_n}^*\sharp\mu_{\tau_n,M})_{n,M\in\mathbb N_+}\) by deriving uniform moment bounds through time averaging. A similar argument was used in~\cite{GH21}, which establishes the existence of a $\Phi^4$ measure on $\sS'(\bR^3)$.

\begin{prop}\label{prop:tightness}
Let $\kappa\in(0,1/3)$. There exists a constant $C\in(0,\infty)$ such that for all $\tau\in\cF$, $M\in \bN_+$, it holds that
\begin{equation}
 \int\|\jmath^*_\tau\phi\|^4_{L_4^{-\kappa}(\bR\times S,w^{1/4})}\,\mu_{\tau,M}(\rd\phi)\leq C.
\end{equation}
\end{prop}
\begin{proof}
On account of Remark~\ref{rem:meaures_cyl_1}, one has
$\mathrm{Law}(\jmath_\tau^*\varPhi_{\tau,M}(t,\Cdot))=\jmath_\tau^*\sharp\mu_{\tau,M}$ for all $t\geq 0$. This implies that
\begin{multline}
\int\|\phi\|^4_{L_4^{-\kappa}(\bR\times S,w^{1/4})}\,(\jmath_\tau^*\sharp\mu_{\tau,M})(\rd\phi)
\\=
\int\|\jmath^*_\tau\phi\|^4_{L_4^{-\kappa}(\bR\times S,w^{1/4})}\,\mu_{\tau,M}(\rd\phi)
=\bE\|\jmath_\tau^*\varPhi_{\tau,M}(t,\Cdot)\|^4_{L_4^{-\kappa}(\bR\times S,w^{1/4})}\,.
\end{multline}
From stationarity, one has $\mathrm{Law}(\jmath_\tau^*X_{\tau, M})=\mathrm{Law}(\jmath_\tau^*Z_{\tau,M}(t,\Cdot))$ for all $t\geq 0$. Thus Lemma~\ref{lem:stochastic_bound_infinite_volume} and
Proposition~\ref{prop:energy_method_plane} imply
\begin{equation}\label{eq: first priori bound}
\,\partial_t\bE\|\jmath_\tau^* \varPsi_{\tau,M}(t,\Cdot)\|^2_{L_2(\bR\times S,w^{1/2})}
 +
\bE \|\jmath_\tau^* \varPsi_{\tau,M}(t,\Cdot)\|^4_{L_4(\bR\times S,w^{1/4})}
\leq C_1 \,
\end{equation}
for some constant $C_1\in(0,\infty)$ independent of $\tau\in\cF$ and $M$ and $t$. For all $T\in(0,\infty)$, the above inequality implies that
\begin{multline}\label{eq:tightness_time_average}
\frac{1}{T}\int_0^T\bE \|\jmath_\tau^* \varPsi_{\tau,M}(t,\Cdot)\|^4_{L_4(\bR\times S,w^{1/4})}\rd t
\\
\leq C_1-\frac{1}{T}\bE\,\|\jmath_\tau^*\varPsi_{\tau,M}(T,\Cdot)\|^2_{L_2(\bR\times S, w^{1/2})}
+\frac{1}{T}\bE\|\jmath_\tau^*\varPsi_{\tau,M}(0,\Cdot)\|^2_{L_2(\bR\times S, w^{1/2})}
\leq C_1 + \frac{C_{\tau,M}}{T}\,,
\end{multline}
where
\begin{multline}
C_{\tau,M}:=\bE\|\jmath_\tau^*\varPsi_{\tau,M}(0,\Cdot)\|^2_{L_2(\bR\times S, w^{1/2})}
=\mathbb E\int_{\mathbb R\times S}
w(s)|\jmath_\tau^*\varPsi_{\tau,M}(0,s,x)|^2\,\rd s\,\rd x\\
=\mathbb E\int_{\Sigma_\tau}
w_\tau(\rz)|\varPsi_{\tau,M}(0,\rz)|^2\,\rd A_\tau(\rz)
\le C_{a,b}\,
\mathbb E\|\varPsi_{\tau,M}(0)\|_{L_2(\Sigma_\tau)}^2<\infty.
\end{multline}
From Remark~\ref{rem:weight_2}, one shows that
\[
C_{a,b}:=
1+2\sum_{k=1}^\infty
\left(1+\frac{a}{4}(k\tau_*)^2\right)^{-b/2}<\infty,
\qquad \tau_*=\frac{\sqrt3}{2},
\]
is independent of \(\tau\in\mathcal F\) and \(M\). From the fact that $\jmath_\tau^*\varPhi_{\tau,M}$ and $\jmath_\tau^*Z_{\tau,M}$ are stationary in time, it follows that
\begin{multline}\label{eq: bound for expectation phi}
\bE\|\jmath_\tau^*\varPhi_{\tau,M}(0, \Cdot)\|^4_{L^{-\kappa}_4(\bR\times S, w^{1/4})}
=
\frac{1}{T}\int_0^T \bE\|\jmath_\tau^*\varPhi_{\tau,M}(t, \Cdot)\|^4_{L^{-\kappa}_4(\bR\times S, w^{1/4})}\rd t
\\
\leq 
{c}\,\bE\|\jmath_\tau^*Z_{\tau,M}(0,\Cdot)\|^4_{L^{-\kappa}_4(\bR\times S, w^{1/4})}+ \frac{c}{T}\int_0^T\bE\|\jmath_\tau^*\Psi_{\tau,M}(t, \Cdot)\|^4_{L^{-\kappa}_4(\bR\times S, w^{1/4})}\rd t\,,
\end{multline}
where $c=2^{3}$. From Lemma~\ref{lem:stochastic_bound_infinite_volume}, there exists $C_2\in(0,\infty)$ such that for all $\tau\in\cF$ and $M\in\bN_+$,
\begin{equation}\label{eq:tightness_free}
 \bE\|\jmath_\tau^*Z_{\tau,M}(0,\Cdot)\|^4_{L^{-\kappa}_4(\bR\times S, w^{1/4})}\leq C_2.
\end{equation}
Using the continuous embedding $L_4(\mathbb R\times S,w^{1/4})
\rightarrow L_4^{-\kappa}(\mathbb R\times S,w^{1/4})$ combined with the preceding bounds yields 
\begin{equation}
 \bE\|\jmath_\tau^*\varPhi_{\tau,M}(0, \Cdot)\|^4_{L^{-\kappa}_4(\bR\times S , w^{1/4})}\leq c\,C_1+c\,C_2 + \frac{c\,C_{\tau,M}}{T}
\end{equation}
for all $T\in(0,\infty)$. Letting $T\to\infty$ concludes the proof.
\end{proof}

\begin{Remark}\label{rem:tightness}
By Lemma~\ref{lem:stochastic_bound_infinite_volume}, the elementary inequality $|a^4-b^4|\leq 2\,|a-b|(|a|^3+|b|^3)$, H\"older's and Minkowski's inequalities, together with the exponential moment bound from Proposition~\ref{prop:uv_limit}, one obtains the convergence
\begin{multline}\label{bound_monet_1}
|\bE\bigl[\|\jmath_\tau^*X_{\tau,M}\|_{L_4^{-\kappa}(\bR\times S,w^{1/4})}^4\exp(-Y_{\tau,M})]-
\bE [
\|\jmath_\tau^*X_\tau\|_{L_4^{-\kappa}(\bR\times S,w^{1/4})}^4\exp(-Y_{\tau,M})] |
\longrightarrow 0
\end{multline}
as $M\to\infty$ for each fixed $\tau\in\mathcal F$. Moreover, Lemmas~\ref{lem:stochastic_bound_infinite_volume} and~\ref{prop:uv_limit} imply uniform integrability of the sequence $(\|\jmath_\tau^*X_\tau\|_{L_4^{-\kappa}(\bR\times S,w^{1/4})}^4\exp(-Y_{\tau,M}))_{M\in\bN_+}$. Since $Y_{\tau,M}\to Y_\tau$ in probability by Lemma~\ref{lem:stochastic_Y}, Vitali's theorem yields
\begin{equation}\label{bound_monet_2}
\bE\,[
\|\jmath_\tau^*X_\tau\|_{L_4^{-\kappa}(\bR\times S,w^{1/4})}^4\exp(-Y_{\tau,M})]
\longrightarrow
\bE\,[\|\jmath_\tau^*X_\tau\|_{L_4^{-\kappa}(\bR\times S,w^{1/4})}^4\exp(-Y_\tau)]
\end{equation}
as $M\to\infty$. Combining Eqs.~\eqref{bound_monet_1} and~\eqref{bound_monet_2} results in
\begin{equation}
\bE\,[
\|\jmath_\tau^*X_{\tau,M}\|_{L_4^{-\kappa}(\bR\times S,w^{1/4})}^4\exp(-Y_{\tau,M})] \longrightarrow
\bE\,[\|\jmath_\tau^*X_\tau\|_{L_4^{-\kappa}(\bR\times S,w^{1/4})}^4\exp(-Y_\tau)]
\end{equation}
as $M$ approaches infinity. From Prop~\ref{prop:uv_limit} with $F=1$ one shows that
$\bE[\exp(-Y_{\tau,M})]$ converges to $\bE[\exp(-Y_\tau)]>0$. Consequently,
\begin{equation}
\int
\|\jmath_\tau^*\phi\|^4_{L_4^{-\kappa}(\bR\times S,w^{1/4})}
\,\mu_\tau(\rd\phi) =
\lim_{M\to\infty}\int\|\jmath_\tau^*\phi\|^4_{L_4^{-\kappa}(\bR\times S,w^{1/4})}\,\mu_{\tau,M}(\rd\phi)
\leq C.
\end{equation}
Note that the embedding ${L^{-\kappa}_4(\bR\times S, w^{1/4})}\to L^{-2\kappa}_4(\bR\times S, w^{1/2})$ is compact by~\cite[Sec.~4.2.3]{EdmundsTriebel}. Consequently,~\cite[Lemma~B.4]{DDJ} implies that the family of measures $(\jmath_\tau^*\sharp\mu_\tau)_{\tau\in\cF}$ is tight on $L^{-2\kappa}_4(\bR\times S, w^{1/2})$. Now Prokhorov's theorem implies that for any sequence \(\tau_n\in\mathcal F\) with \(\operatorname{Im}\tau_n\to\infty\), there exists a subsequence \((\tau_{n_k})\) such that for every bounded and continuous functional $F\,:\,L^{-2\kappa}_4(\bR\times S, w^{1/2})\to\bR$, it holds that
\[
\mu(F)
=\lim_{k\to\infty}
(\jmath_{\tau_{n_k}}^*\sharp\mu_{\tau_{n_k}})(F)
=\lim_{k\to\infty}\lim_{M\to\infty}
(\jmath_{\tau_{n_k}}^*\sharp\mu_{\tau_{n_k},M})(F)
=\lim_{k\to\infty}\lim_{M\to\infty}
\mu_{\tau_{n_k},M}(F\circ\jmath_{\tau_{n_k}}^*).
\]
\end{Remark}


\appendix

\section{Function space}
Consider the flat torus $
\Sigma_\tau=\mathbb C/(\mathbb Z+\tau\mathbb Z)$, where $\tau=\tau_1+i\tau_2\in\cF$,
with coordinates $\rz=u\tau+v,\quad (u,v)\in (0,1]^2$ and the volume form $\rd A_{\tau}=\tau_2\,\rd u\,\rd v$. 
\begin{dfn}
We define the injectivity radius of $\Sigma_\tau$ by
\[
\operatorname{inj}(\Sigma_\tau)
=\frac12\min_{(m,n)\in\mathbb Z^2\setminus\{(0,0)\}}|m+n\tau|.
\]
For \(\tau\in\mathcal F\), we have $
|\tau_1|\leq\frac12$, $|\tau|\ge1$, $\tau_2\ge\frac{\sqrt3}{2}$. This implies that $
\mathrm{inj}(\Sigma_\tau)=\frac12$ for every $\tau\in\mathcal F$.

\end{dfn}
\begin{Remark}
Let $\tau\in\cF$. Using the metric $g_\tau=|\rd z|^2 = |\tau|^2\,\rd u^2 + 2\tau_1\,\rd u\,\rd v + \rd v^2$, the Laplacian on $\Sigma_\tau$ is given by $\Delta_{\tau} =\tau_2^{-2}\,
\left(\partial_u^2 +|\tau|^2\partial_v^2 -2\tau_1 \partial_u\partial_v\right)$.
Let $f_{m,n}(u,v)=\re^{2\pi i(mu+nv)}$. One obtains 
\begin{equation}\label{eq:lap_eig}
\Delta_{\tau}  f_{m,n}
=-\frac{4\pi^2}{\tau_2^2} |m-\tau n|^2 f_{m,n}:=-\lambda_{m,n}(\tau)f_{m,n}.
\end{equation}
Since the periodic delta on \([0,1)^2\) has Fourier expansion
$\delta(u)\delta(v)
=\sum_{(m,n)\in\mathbb Z^2}\re^{2\pi i(mu+nv)}$, understood in the distributional sense on $\Sigma_\tau$,
we obtain
\[
\delta_\tau(\rz)=\frac{1}{\tau_2}
\sum_{(m,n)\in\mathbb Z^2}\re^{2\pi i(mu+nv)},
\]
which satisfies $\int_{\Sigma_\tau}\delta_\tau(\rz)\,f(\rz)\,\rd A_\tau(\rz)=f(0)$.
\end{Remark}
In what follows, we consider the polynomial weight \(w(s)=(1+as^2)^{-b/2}\), where \(b>2\) and \(a\in(0,1)\) is sufficiently small, see Remark~\ref{rem:weight_2}.
\begin{dfn}\label{def:weighted_norm}
For \(1\leq p<\infty\), we define
\begin{equation}
\| f\|_{L^{p}(\Sigma_{\tau})} :=\Big(\int_{\Sigma_{\tau}} |f(\rz)|^{p} \rd A_\tau(\rz)\Big)^{1/p},
\quad
L^{p}(\Sigma_{\tau}) := \Big\{ f \in  \cD'(\Sigma_{\tau}): \| f\|_{L^{p}(\Sigma_{\tau})} < \infty \Big\}.
\end{equation}
For \(\alpha\in\mathbb R\) and $p\in[1,\infty)$, define the Sobolev norm
\[
\|f\|_{L_p^\alpha(\Sigma_\tau)}
=
\|(1-\Delta_\tau)^{\alpha/2}f\|_{L_p(\Sigma_\tau)}.
\]
Similarly, on $\bR\times S$ we define
\[
\|f\|_{L_p^\alpha(\bR\times S,w^{1/p})}
=
\|(1-\Delta_{\mathrm{cyl}})^{\alpha/2}f\|_{L_p(\bR\times S,w^{1/p})},
\]
where $(-\Delta_{\mathrm{cyl}})$ is the Laplacian on $\bR\times S$ given in Eq.~\eqref{eq:intertwining}.
\end{dfn}

\begin{lem}\label{lem:lp-l2}
Let $p\in[2,\infty)$, and $\alpha=1-2/p$. There exists $C\in(0,\infty)$ such that $\|f\|_{L_p(\Sigma_\tau)}\leq C\,\|f\|_{L_2^\alpha(\Sigma_\tau)}$ for all $f\in L_2^\alpha(\Sigma_\tau)$ and $\tau\in\cF$.
\end{lem}
\begin{Remark}\label{rem:bessel_sobolev_psi}
Let $\tau\in\cF$ and $p\in[2,\infty)$. By Lemma~\ref{lem:lp-l2}, there exist $C,C' \in (0,\infty)$ such that for all $\Psi\in L_2^{2-2/p}(\Sigma_\tau)$, one has
\begin{equation}
\|\nabla_\tau\Psi\|_{L_p(\Sigma_\tau)}
\leq C\,\|\nabla_\tau\Psi\|_{L_2^{1-2/p}(\Sigma_\tau)}
\leq C\,\|(1-\Delta_\tau)^{1/2}\, \Psi\|_{L_2^{1-2/p}(\Sigma_\tau)}
\leq C'\,\|\Psi\|_{L_2^{2-2/p}(\Sigma_\tau)}.
\end{equation}
\end{Remark}

\begin{Remark}\label{rem:holder_weight}
For $\alpha\in\mathbb R$ and $p,q\in(1,\infty)$ with $1/p+1/q=1$, the following generalized H\"older inequality holds
\[
|\langle f,g\rangle_{L_2(\mathbb R\times S,w^{1/2})}|\le C_{\alpha,p,w}\,
\|f\|_{L_p^\alpha(\bR\times S,w^{1/p})}\,
\|g\|_{L_q^{-\alpha}(\bR\times S,w^{1/q})},
\]
where $C_{\alpha,p,w}\in(0,\infty)$ depends on $\alpha,p$, and $w$.
\end{Remark}

\begin{lem}\label{lem:Z_Psi_4}
Let $\delta\in(0,\infty)$ and $\kappa\in(0,1/3)$. 
Then there exist $C\in(0,\infty)$ and $p\in[1,\infty)$ such that for all \mbox{$m\in\{1,2,3\}$} and $\Psi\in L_2^1(\bR\times S,w^{1/2}) \cap L_4(\bR\times S,w^{1/4})$, $Z\in L^{-\kappa}_p(\bR\times S,w^{1/p})$  it holds
\begin{equation}
 |\langle Z,\Psi^m \rangle_{L_2(\bR\times S ,w^{1/2})}| 
 \leq
 C\,\|Z\|_{L^{-\kappa}_p(\bR\times S,w^{1/p})}^p+
 \delta\,\|\nabla\Psi\|_{L_2(\bR\times S,w^{1/2})}^2+
 \delta\,\|\Psi\|_{L_4(\bR\times S,w^{1/4})}^4+\delta.
\end{equation}
\end{lem}
\begin{proof}
The proof follows from Remark~\ref{rem:holder_weight} by adapting the argument of~\cite[Lemma.~A.8]{DDJ}.
\end{proof}
\begin{Remark}\label{rem:spec_heat_sig}
Let \(d_\tau\) be the geodesic distance on \(\Sigma_\tau=\bC/(\bZ+\tau\bZ)\) induced by \(g_\tau=|\rd \rz|^2\):
\[
d_\tau([\rz],[\rz'])
=\min_{m,n\in\mathbb Z}|\rz-\rz'+m+n\tau|.
\]
For $t>0$ and $\rz=\tau u+v\in\Sigma_\tau$, let $H_\tau(t,\rz)$ denote the heat kernel associated with $(-\Delta_\tau)$
\begin{equation}\label{eq:heat_periodic_1}
H_\tau(t,\rz)
=\frac1{4\pi t}
\sum_{\ell\in\mathbb Z+\tau\mathbb Z}
\exp\left(-\frac{|z+\ell|^2}{4t}\right)
=\frac1{\tau_2}\sum_{m,n\in\mathbb Z}
\re^{-t\lambda_{m,n}(\tau)}
\re^{2\pi i(mu+nv)},
\end{equation}
where $\lambda_{m,n}(\tau)$ is given in Eq.~\eqref{eq:lap_eig}. For every \(\beta>1\), one has
\begin{equation}\label{eq:spec_heat_sig_1}
\sup_{\tau\in\mathcal F}
\frac1{\tau_2}
\sum_{m,n\in\mathbb Z}
(1+\lambda_{m,n}(\tau))^{-\beta}
<\infty.
\end{equation}
Note that on $\cF$ every nonzero lattice vector has length at least \(1\). Consequently, there is $C\in (0,\infty)$ such that for all $\tau\in\cF$ one has
\begin{equation}\label{eq:heat_periodic_2}
H_\tau(t,\rz)\le
\frac C t\exp\left(-\frac{d_\tau(\rz,0)^2}{8t}\right)
\qquad 0<t\le1.
\end{equation}
Let $M\in(1,\infty)$ and consider $
G_{\tau,M}=(1-\Delta_\tau)^{-1}(1-M^{-2}\Delta_\tau)^{-2}$. For $a>0$, one has $(a)^{-1}=\int_0^\infty \re^{-sa}\,\rd s$, $(a)^{-2}=\int_0^\infty s\, \re^{-sa}\,\rd s$. This implies
\begin{equation}\label{eq:heat_reg_1}
G_{\tau,M}(\rz)
=\int_0^\infty\int_0^\infty
t \,\re^{-t}\re^{-s}
H_\tau\left(s+\frac{t}{M^2},\rz\right)\,\rd s\,\rd t
=\int_0^\infty K_M(u)\,H_\tau(u,\rz)\,\rd u,
\end{equation}
where \(u=s+t/M^2\) and
$
K_M(u)=\re^{-u}\int_0^{M^2u}
t\,\re^{-(1-M^{-2})t}\,\rd t$. Moreover, $ 0\le K_M(u)\leq C \,\re^{-u/2}$, for all $M\geq 1$. Note that \(G_{\tau,M}\) and \(G_\tau\) are functions of $(-\Delta_\tau)$, this implies that both are simultaneously diagonalizable. Therefore, in operator sense
\[
0\leq G_{\tau,M}\leq G_\tau,
\]
where $G_\tau(\rz)=\int_0^\infty \re^{-t}H_\tau(t,\rz)\,\rd t$. Eq.~\eqref{eq:heat_reg_1} and the bound on $K_M$ imply that
\[
0\le G_{\tau,M}(\rz)
\le C\int_0^\infty \,\re^{-u/2}H_\tau(u,\rz)\,\rd u.
\]
Using Eqs.~\eqref{eq:heat_periodic_1}~and~\eqref{eq:heat_periodic_2} and setting \(r=d_\tau(\rz,0)\), we obtain, for \(0<r\le1/2\),
\[
G_{\tau,M}(\rz)
\le C+
C\int_0^1\frac1u
\exp\left(-\frac{r^2}{8u}\right)\,\rd u
\le C(1+|\log r|).
\]
For \(r>1/2\), the same estimates give a uniform constant. Hence
\begin{equation}\label{eq:large_kernel}
0\leq G_{\tau,M}(\rz)
\leq C\left(1+|\log\bigl(d_\tau(\rz,0)\wedge1\bigr)|
\right)
\end{equation}
holds for every \(\tau\in\mathcal F\), \(M>1\), and \(\rz\ne0\), with \(C\) independent of \(\tau\in\cF\) and \(M\in \bN_+\).
\end{Remark}
For general heat-kernel theory on Riemannian manifolds, see~\cite{Va03}~and~\cite[Ch.~7]{Gri09}.
\section{Uniform renormalization}
\begin{proof}[Proof of Lemma~\ref{lem:counter-term}]\label{proof:counteterm}
Using the identity
\[
\frac1{(1+s)(1+s/M^2)^2}
=
\frac{M^4}{(M^2-1)^2} \left[
\frac1{1+s}
-\frac1{M^2+s}
-\frac{M^2-1}{(M^2+s)^2}\right],
\]
one writes
\[
G_{\tau,M}(0)
=\frac{1}{\tau_2}\frac{M^4}{(M^2-1)^2}
\sum_{(m,n)\in\mathbb Z^2}
\left[
\frac{1}{1+\lambda_{m,n}(\tau)}
-\frac{1}{M^2+\lambda_{m,n}(\tau)}
-\frac{M^2-1}{(M^2+\lambda_{m,n}(\tau))^2}
\right].
\]
From the heat kernel for \(s\ge0\) one obtains
\[
\frac{1}{(1+s)(1+s/M^2)^2}
=
\frac{M^4}{(M^2-1)^2}
\int_0^\infty
\left[
\re^{-t}-\re^{-M^2t}-(M^2-1)\,t\,\re^{-M^2t}
\right]\re^{-ts}\,\rd t.
\]
For \(k=(m,n)^T\in\mathbb Z^2\), we have $
\lambda_{m,n}(\tau)=4\pi^2\, k^T\,g_\tau^{-1}k$. Furthermore, Poisson summation~\cite[Eq.~(4.4)]{Va03}~or~\cite[Sec.~3.1.2]{Eli95}
$$
\sum_{k\in\mathbb Z^2}
\re^{-4\pi^2\,t\, k^Tg_\tau^{-1}k}=
\frac{\tau_2}{4\pi t}
\sum_{k\in\mathbb Z^2}\re^{-k^Tg_\tau k/(4t)}
$$
implies that
\[
\sum_{m,n\in\bZ}
\re^{-t\lambda_{m,n}(\tau)}
=
\frac{\tau_2}{4\pi t}
\sum_{p,q\in\bZ}
\exp\left(-\frac{|p+q\tau|^2}{4t}\right).
\]
Therefore,
\[
G_{\tau,M}(0)
=\frac{M^4}{(M^2-1)^2}\frac1{4\pi}
\sum_{p,q\in\mathbb Z}
\int_0^\infty
\frac{\re^{-t}-\re^{-M^2t}-(M^2-1)\,t\,\re^{-M^2t}}{t}
\exp\left(-\frac{|p+q\tau|^2}{4t}\right)\,\rd t.
\]
For the zero lattice vector, the integral is
\[
\int_0^\infty
\frac{\re^{-t}-\re^{-M^2t}-(M^2-1)\,t\,\re^{-M^2t}}{t}\,\rd t
=
2\log M-\frac{M^2-1}{M^2}.
\]
For \(r>0\), using
\[
\int_0^\infty\frac{\re^{-at}}{t}\re^{-r^2/(4t)}\,\rd t
=2K_0(r\sqrt a),
\qquad
\int_0^\infty\re^{-at}\re^{-r^2/(4t)}\,\rd t
=\frac{r}{\sqrt a}K_1(r\sqrt a),
\]
we obtain 
\begin{multline}
G_{\tau,M}(0)
=\frac{M^4}{(M^2-1)^2}\frac1{4\pi}
\bigg[
2\log M-\frac{M^2-1}{M^2}
\\
+\sum_{(p,q)\ne(0,0)}
\left\{
2\bigl[K_0(|p+q\tau|)-K_0(M|p+q\tau|)\bigr]
-\frac{M^2-1}{M}|p+q\tau|K_1(M|p+q\tau|)
\right\}
\bigg].
\end{multline}
Consequently, for each fixed \(\tau\in\cF\),
\[
G_{\tau,M}(0)-\frac{\log M}{2\pi}
=
\frac1{2\pi}\sum_{(p,q)\ne(0,0)}K_0(|p+q\tau|)
-\frac1{4\pi}
+O_\tau\!\left(\frac{\log M}{M^2}\right)
+O_\tau\!\left(M^{1/2}\,\re^{-c_\tau M}\right),
\]
where $c_\tau=\min_{(p,q)\ne(0,0)}|p+q\tau|>0$. Hence
\[
\lim_{M\to\infty}|G_{\tau,M}(0)-\frac{\log M}{2\pi}|
=|
\frac1{2\pi}\sum_{(p,q)\ne(0,0)}K_0(|p+q\tau|)
-\frac1{4\pi}|.
\]
For each fixed \(\tau\in\cF\) as \(M\to\infty\) it holds
\[
\lim_{M\to\infty}|G_{\tau,M}(0)-\frac{\log M}{2\pi}|
\leq 
\frac1{2\pi}\sum_{(p,q)\ne(0,0)}
K_0(|p+q\tau|)+\frac{1}{4\pi},
\]
since $K_0(r)=O(r^{-1/2}\re^{-r})$ as $r\to\infty$.
Using the fact that the shortest nonzero lattice length on $\cF$ is bounded below by $1$, combined with the decaying property of $K_0$ yields
$
\sup_{\tau\in\cF}
\sum_{(p,q)\ne(0,0)}K_0(|p+q\tau|)<\infty$. This finishes the proof.
\end{proof}

\begin{lem}\label{lem:hat_counteterm}
There is $C\in (0,\infty)$ such that
\begin{equation}
\sup_{\tau\in\cF}\lim_{M\to\infty} |\hat c_{\tau,M}-\frac{1}{2\pi}\log M|<C\,.
\end{equation}
\end{lem}
\begin{proof}
Using  the periodization formula for $G_\tau$ one has
$
G_\tau(z,w)=(2\pi)^{-1}\sum_{m,n\in\bZ}K_0(|z-w+m+n\tau|)$. Thus
\begin{equation}
\hat c_{\tau,M}:=
\frac1{2\pi}
\sum_{m,n\in\bZ}
\int_{\bR^2}\int_{\bR^2}
\rho(u)\,\rho(v)\,K_0\,(|m+n\tau+\frac{u-v}{M}|)\,\rd u\,\rd v
:= I_M+R_{\tau,M},
\end{equation}
where $I_M$ and $R_{\tau,M}$ denote the zero mode and the remainder contribution, respectively. By the support property of $\rho$ combined with the asymptotic behavior of $K_0$ near the origin, one shows that $
\lim_{M\to\infty}\,(I_M-\frac{\log M}{2\pi})
=C_\rho$, where $C_\rho\in (0,\infty)$ is independent of $\tau\in\cF$. Using the fact that the shortest nonzero lattice length on $\mathcal F$ is bounded below by $1$, one verifies by dominated convergence, that
$
\lim_{M\to\infty}\,R_{\tau,M}
=(2\pi)^{-1} \sum_{\lambda\in\Lambda_\tau\setminus\{0\}}K_0(|\lambda|)$. The above results culminate in
\begin{equation}
\lim_{M\to\infty}
\left(\hat c_{\tau,M}
-\frac{\log M}{2\pi}\right)
=C_\rho+
\frac1{2\pi}
\sum_{\lambda\in\Lambda_\tau\setminus\{0\}}
K_0(|\lambda|).
\end{equation}
Invoking the fact that $\min_{(m,n)\in\bZ^2\setminus\{(0,0)\}}|m+n\tau|=1$ and the decay property of $K_0$, one shows that the second term on the RHS of the last line is uniformly bounded in $\tau\in\cF$ and this finishes the proof.
\end{proof}
\section{Stochastic estimates}
\begin{lem}\label{lem:sigma_1}
For every $\kappa\in(0,\infty)$, $M\in\bN_+$ there exists $C\in(0,\infty)$ such that for all $\tau\in\cF$ it holds
$\bE\|X_{\tau,M}\|_{L^{2-\kappa}_2(\Sigma_\tau)}^2 \leq \, \tau_2\,C^2$.
\end{lem}
\begin{proof}
By the spectral decomposition,
\[
\mathbb E\|X_{\tau,M}\|_{L_2^{2-\kappa}(\Sigma_\tau)}^2
=\sum_{m,n\in\mathbb Z}
\frac{(1+\lambda_{m,n}(\tau))^{1-\kappa}}
{(1+M^{-2}\lambda_{m,n}(\tau))^2}
\le M^4\sum_{m,n\in\mathbb Z}
(1+\lambda_{m,n}(\tau))^{-1-\kappa}.
\]
The last bound follows from Eq.~\eqref{eq:spec_heat_sig_1}.
\end{proof}

\begin{lem}\label{lem:stochastic_X_uniform}
For every $\kappa\in(0,\infty)$, $\delta\in[0,2]$, there exists $C\in(0,\infty)$ such that for all $\tau\in\cF$ and $M\in\bN_+$ it holds that
\begin{enumerate}
 \item[(A)] $\bE\|X_\tau\|_{L^{-\kappa}_2(\Sigma_\tau)}^2 
  \leq \tau_2\,C^2$,
 \item[(B)] $\bE\|X_\tau- X_{\tau,M}\|_{L_2^{-\kappa-\delta}(\Sigma_\tau)}^2 
 \leq \tau_2\,C^2\, M^{-2\delta}$.
\end{enumerate}
\end{lem}
\begin{proof}
By definition and Eq.~\eqref{eq:spec_heat_sig_1} one obtains
$$
\mathbb E\|X_\tau\|_{L_2^{-\kappa}(\Sigma_\tau)}^2
=\sum_{m,n\in\mathbb Z}
(1+\lambda_{m,n}(\tau))^{-1-\kappa}\leq C^2\,\tau_2\,.$$
To prove Item~(B), one writes
\begin{equation}
\begin{aligned}
\mathbb E\|X_\tau-X_{\tau,M}\|_{L_2^{-\kappa-\delta}(\Sigma_\tau)}^2
&=\sum_{m,n\in\mathbb Z}
\frac{1}{(1+\lambda_{m,n}(\tau))^{1+\kappa+\delta}}
\left(\frac{\lambda_{m,n}(\tau)}{M^2+\lambda_{m,n}(\tau)}\right)^2
\\&\le M^{-2\delta}
\sum_{m,n}(1+\lambda_{m,n}(\tau))^{-1-\kappa}
\leq C^2\,\tau_2\,M^{-2\delta},
\end{aligned}
\end{equation}
where we used $(\frac{\lambda}{M^2+\lambda})^2\leq M^{-2\delta}(1+\lambda)^\delta$ for $0\leq \delta\leq 2$. The last bound follows from Eq.~\eqref{eq:spec_heat_sig_1}.
\end{proof}

\begin{lem}\label{lem:stochastic_Y}
Let $\tau\in\cF$. There exists a real-valued random variable $Y_\tau$ and $C\in(0,\infty)$ such that for all $M\in\bN_+$ it holds: (A)\,$\bE Y_\tau^2\leq C^2$, and (B)\,$\bE (Y_\tau-Y_{\tau,M})^2\leq C^2\,M^{-1/4}$.
\end{lem}
\begin{proof}
We establish the required local regularity estimates for the regularized covariance. For all $N,M\in\bN_+$, define
$G_{\tau,N,M}:=(1-M^{-2}\Delta_\tau)^{-1}\,(1-\Delta_\tau)^{-1}\,(1-N^{-2}\Delta_\tau)^{-1}$. Let $\tau\in\cF$. Using the translational invariance of $G_{\tau,M}$, Lemma~\ref{lem:lp-l2} combined with Lemma~\ref{lem:counter-term} and the choice of the counterterm $c_{\tau,M}$, one shows that there exists $C\in (0,\infty)$ such that for all $N,M\in\bN_+$ with $N\wedge M=N$, it holds
$$\sup_{\rz\in\Sigma_\tau}\|G_{\tau,N,N}(\rz-\Cdot)\|_{L_4(\Sigma_\tau)} \leq C,\quad \|G_{\tau,N,N}- G_{\tau,N,M}\|_{L_4(\Sigma_\tau^2)}\leq C\,(N^{-1/4}).$$
Now using the covariance identity for Wick powers, i.e., $
\bE\,[
X_{\tau,M}^{:m:}\otimes X_{\tau,M}^{:m:}]=m!\,G_{\tau,M}^m$ completes the proofs of~(A) and~(B).
\end{proof}

\begin{lem}\label{lem:stochastic_bound_infinite_volume}
Let $m\in\bN_+$, $p\in[1,\infty)$, $\kappa\in(0,\infty)$. There exists $C\in(0,\infty)$ such that for all $\tau\in\cF$ and $M\in\bN_+$, it holds that
\begin{itemize}
\item[(A)]
$ \bE\|\jmath_\tau^* X^{:m:}_{\tau,M}\|_{L^{-\kappa}_p(\bR\times S,w^{1/p})}^p \leq C$,
\item[(B)]
$
\lim_{M\to\infty}\mathbb E\|\jmath_\tau^*(X_\tau-X_{\tau,M})\|_{L_p^{-\kappa}(\mathbb R\times S,w^{1/p})}^p
=0$.
\end{itemize}
\end{lem}

\begin{proof}
Let $K_{\tau,\kappa}$ denote the integral kernel of $(1-\Delta_\tau)^{-\kappa}$. By Wick's and Fubini's theorem, one obtains
\[
\bE\,|
(1-\Delta_\tau)^{-\kappa/2}
X_{\tau,M}^{:m:}(\rz)|^2=m!\int_{\Sigma_\tau}
K_{\tau,\kappa}(\rz')\,G_{\tau,M}(\rz')^m\,\rd A_\tau(\rz').
\]
Eq.~\eqref{eq:large_kernel} implies that $
G_{\tau,M}(\rz)^m
\le C^m\,(1+|\log(d_\tau(\rz,0)\wedge1)|)^m$ uniformly in $\tau\in\cF$.  One also finds an upper bound on $K_{\tau,\kappa}$ using Eq.~\eqref{eq:heat_periodic_2}. Consequently,
\[
\sup_{\tau\in\mathcal F}\sup_{M\ge1}
\int_{\Sigma_\tau}
K_{\tau,\kappa}(\rz)\,G_{\tau,M}(\rz)^m\,\rd A_\tau(\rz)
\le c
\]
for some $c\in (0,\infty)$. Hence, there is $c'\in (0,\infty)$ such that
\[
\sup_{\tau,M,z}
\mathbb E\,|
(1-\Delta_\tau)^{-\kappa/2}
X_{\tau,M}^{:m:}(\rz)|^2
\le c'.
\]
By hypercontractivity on the \(m\)-th homogeneous Wiener chaos, for \(p\ge2\), and Jensen's inequality for \(1\le p<2\),
\[
\sup_{\tau,M,z}
\mathbb E\,|
(1-\Delta_\tau)^{-\kappa/2}X_{\tau,M}^{:m:}(\rz)|^p\le c_2,
\]
where $c_2\in (0,\infty)$. Using Eq.~\eqref{eq:intertwining} and Tonelli's theorem, one obtains
\[
\mathbb E\|\jmath_\tau^*X_{\tau,M}^{:m:}\|_
{L_p^{-\kappa}(\mathbb R\times S,w^{1/p})}^p=\int_{\mathbb R\times S}\,w(s)
\mathbb E\,|
\jmath_\tau^*(1-\Delta_\tau)^{-\kappa/2}
X_{\tau,M}^{:m:}(s,x)|^p\,\rd s\,\rd x
\le c_2\,\int_{\bR}w(s)\,\rd s
\le C.
\]
This finishes the proof of item~(A). To prove item~(B), one uses the spectral decomposition,
\[\mathbb E\,|
(1-\Delta_\tau)^{-\kappa/2}
(X_\tau-X_{\tau,M})(\rz)|^2
=\frac1{\tau_2}\sum_{k,l\in\mathbb Z}
\frac1{(1+\lambda_{k,l}(\tau))^{1+\kappa}}(
1-\frac1{1+\lambda_{k,l}(\tau)/M^2})^2.
\]
For \(0\leq \delta\leq 2\), one has $(1-\frac1{1+\lambda/M^2})^2
=(\frac{\lambda}{M^2+\lambda})^2\le M^{-2\delta}\lambda^\delta$.
Therefore, for some $c'\in(0,\infty)$
\[\mathbb E\,|
(1-\Delta_\tau)^{-\kappa/2}
(X_\tau-X_{\tau,M})(\rz)|^2\le \frac{M^{-2\delta}}{\tau_2}
\sum_{k,l\in\mathbb Z}
(1+\lambda_{k,l}(\tau))^{-1-\kappa+\delta}
\le c'\,M^{-2\delta},
\]
where the last inequality follows from $\delta<\kappa$. Since the difference is Gaussian, one has
\[
\mathbb E\,|
(1-\Delta_\tau)^{-\kappa/2}
(X_\tau-X_{\tau,M})(\rz)|^p
\le c''\,M^{-p\delta}.
\]
Using the intertwining property of $\jmath_\tau$, stationarity, and the integrability of $w$, one deduces
\[
\mathbb E\|\jmath_\tau^*(X_\tau-X_{\tau,M})\|_
{L_p^{-\kappa}(\mathbb R\times S,w^{1/p})}^p
\le c''\,M^{-p\delta}
\int_{\mathbb R\times S}\,w(s)\,\rd s\,\rd x
\le C\,M^{-p\delta}.
\]
This proves (B) and completes the statement.
\end{proof}


\section*{Acknowledgments}
During the final stages of manuscript preparation, ChatGPT was used exclusively for English-language refinement, mathematical proofreading, and bibliographic assistance, including identifying the reference~\cite{Eli95}.


\end{document}